\documentclass[11pt]{article}
\usepackage{amsmath,amsfonts,amssymb,eepic,graphicx,epsfig,epic}
\newcommand{\sepAuthor}{0.6in}
\newcommand{\sepAbstract}{0.6in}
\newcommand{\skipKeywords}{50pt}
\long\def\mytitlepage#1#2#3#4{
        \thispagestyle{empty}
        \begin{center}
        {\Large\bf #1}

        \vspace{\sepAuthor}
        #2\\
        \medskip

        \vspace{\sepAbstract}
        {\Large Abstract}
        \end{center}

        \noindent{#3}
        \vskip\skipKeywords

        \noindent{#4}
        \clearpage
        }
\usepackage{amsthm}
\theoremstyle{plain}
\newtheorem{theorem}{Theorem}[section]
\newtheorem{lemma}[theorem]{Lemma}

\newtheorem{proposition}[theorem]{Proposition}

\theoremstyle{definition}
\newtheorem{definition}[theorem]{Definition}

\newtheorem{algorithm}{Algorithm}
\theoremstyle{remark}

\def\squareforqed{\hbox{\rlap{$\sqcap$}$\sqcup$}}
\def\qed{\ifmmode\squareforqed\else{\unskip\nobreak\hfil
\penalty50\hskip1em\null\nobreak\hfil\squareforqed
\parfillskip=0pt\finalhyphendemerits=0\endgraf}\fi}

\newenvironment{proofof}[1]{\begin{trivlist}%
\item[]{\flushleft\em Proof of #1. }}
{\end{trivlist}}

\newcommand{\bstr}[1]{\mbox{$\{0, 1\}^{#1}$}}

\newcommand{\ket}[1]{\mbox{$\left|{#1}\right\rangle$}}
\newcommand{\bra}[1]{\mbox{$\left\langle{}{#1}\right|$}}

\newcommand{\ub}[1]{\ensuremath{O\left({#1}\right)}}

\newcommand{\inner}[2]{\mbox{$\left\langle #1 \right|\left. #2 \right\rangle$}}

\newcommand{\op}[1]{{\mathsf{#1}}}
\newcommand{\opH}{\ensuremath{\op{H}}}
\newcommand{\opId}{\ensuremath{\op{I}}}

\newcommand{\opSet}[1]{\ensuremath{\mathbf{#1}}}

\newcommand{\set}[1]{\ensuremath{\left\{ {#1} \right\}}}

\newcommand{\norm}[1]{\ensuremath{\left\|{#1}\right\|}}

\newcommand{\SHilbert}[1]{\ensuremath{\mathcal{H}^{#1}}}
\newcommand{\Hspace}{\ensuremath{\mathcal{H}}}

\newcommand{\Span}{\ensuremath{\mathrm{span}}}

\newcommand{\txtCNOT}{\textrm{{}CNOT}}
\newcommand{\txtToffoli}{\textrm{{}Toffoli}}
\newcommand{\txtHadamard}{\textrm{{}Hadamard}}
\newcommand{\comments}[1]{}
\newcommand{\Ctrl}[2][]{\ensuremath{\Lambda^{#1}(#2)}}
\newcommand{\CtrlN}[2][]{\ensuremath{\bar{\Lambda}^{#1}(#2)}}
\newcommand{\define}{\ensuremath{:=}}

\newcommand{\opHadamard}{\ensuremath{H}}
\newcommand{\inverse}[1]{\ensuremath{\frac{1}{#1}}}
\newcommand{\tensor}[2]{\ensuremath{{#1}^{\otimes #2}}}
\newcommand{\opCNOT}{\ensuremath{\Ctrl{\sigma^x}}}
\newcommand{\opToffoli}{\ensuremath{\Ctrl[2]{\sigma^x}}}
\newcommand{\sCopies}[2]{\ensuremath{\tensor{{#1}}{#2}}}

\newcommand{\AppOpWHA}{\tilde{\op{W}}_{\alpha/2}}
\newcommand{\AppSigmaZ}{\tilde{\sigma}^z}
\newcommand{\AppY}{\ket{\tilde{1}}}
\newcommand{\ConstCos}{\delta'_\theta}
\newcommand{\ConstSin}{\delta_\theta}
\newcommand{\ConstIErr}{\inverse{\epsilon}}
\newcommand{\ConstLIErr}{\log\inverse{\epsilon}}

\newcommand{\HalfAlpha}{\ket{\phi_{\alpha/2}}}
\newcommand{\LogHA}{\ket{\hat{\phi}_{\alpha/2}}}
\newcommand{\LogX}{\ket{\hat{0}}}
\newcommand{\LogY}{\ket{\hat{1}}}

\newcommand{\Mixed}{\ket{\tilde{\psi}_3}}
\newcommand{\OpAppSigmaZ}{\tilde{\sigma}^z}

\newcommand{\OpT}{\op{T}_{\theta}}
\newcommand{\opT}{\OpT}
\newcommand{\OpTK}{\op{T}_{\theta}^{\otimes k}}
\newcommand{\OpU}{\op{U}_\theta}
\newcommand{\OpUI}{\op{U}_{-\theta}}

\newcommand{\OpUA}{\op{U}_\alpha}
\newcommand{\OpWA}{\op{W}_\alpha}
\newcommand{\OpWHA}{\op{W}_{\alpha/2}}
\newcommand{\PhaseAncila}[1][k]{\ket{\Phi_{#1}}}
\newcommand{\ReflectY}{\bar{\Lambda}^{2k}(\sigma^z)}

\newcommand{\SHA}{\ket{\phi_{\alpha/2}}}

\newcommand{\RealB}{\ensuremath{\mathfrak{B}}}
\begin{document}
\mytitlepage{
Both Toffoli and Controlled-NOT need little help to do universal
quantum computation
\footnote{This work was supported in part by 
NSF Grant EIA-0086038, NSF Grant 0049092, and 
The Charles Lee Powell Foundation.
}} 
{\large{Yaoyun Shi}\\
\vspace{1ex}
{Computer Science Department and\\
Institute for Quantum Information\\
California Institute of Technology\\
Pasadena, CA 91125\\
E-mail: shiyy@cs.caltech.edu
}}
{What additional gates are needed
for a set of {\em classical} universal gates to
do {\em universal  quantum} computation?
We answer this question by proving that
\emph{any} single-qubit real gate suffices,
except those that preserve the
computational basis. 

The result of
Gottesman and Knill~\cite{Gottesman98} implies
that any quantum circuit involving 
only the Controlled-NOT and Hadamard
gates can be efficiently simulated by a classical circuit.
In contrast, we prove that Controlled-NOT plus any
single-qubit real gate that does not preserve the computational
basis and is not Hadamard (or its alike) are universal
for quantum computing.

Previously only
a ``generic'' gate, namely a rotation
by an angle incommensurate with $\pi$,
is known to be sufficient in both problems, if only one 
single-qubit gate is added.
}
\section{Introduction}\label{sec:intro}
A set of quantum gates $G$ (also called a {\em basis})
is said to be {\em universal
for quantum computation} if any unitary operator can
be approximated with arbitrary precision by a circuit
involving only those gates (called a $G$-circuit). Since
complex numbers do not help in quantum computation,
we also call a set of real gates
universal if it approximates arbitrary real orthogonal
operators.

Which set of gates are universal for quantum
computation?
This basic question is important both
in  understanding the power of quantum computing
and in the physical implementations of quantum computers, 
and has been studied extensively.
Examples of universal bases are:
(1) Toffoli, Hadamard, and $\frac{\pi}{4}$-gate,
due to Kitaev~\cite{Kitaev97} (2) 
\txtCNOT, \txtHadamard, and $\frac{\pi}{8}$-gate,
due to Boykin, Mor, Pulver, Roychowdhury, and Vatan~\cite{BoykinMPRV00},
and (3)
\txtCNOT\ plus the set of all single-qubit gate,
due to Barenco, Bennett, Cleve,
DiVincenzo, Margolus, Shor, Sleator, Smolin, and 
Weinfurter~\cite{Barenco+95}.

Another basic question in understanding quantum computation is, 
Where does the power of quantum computing come from?
Motivated by this question, we rephrase the universality 
question as follows:
Suppose a set of gates $G$ already 
contains universal classical gates,
and thus can do universal classical computation,
what additional quantum gate(s) does it need to do universal quantum
computation? Are there some gates that are 
{\em more ``quantum''} than some others in
bringing more computational power?

Without loss of generality, we assume that $G$
contains the Toffoli gate, since it is universal for
classical computation. The above three examples of
universal bases provide some answers to this question.
It is clear that we need at least one additional
gate that does not preserve the computational basis.
Let us call such a gate \emph{basis-changing}.
Our main result is that essentially the basis-changing
condition is the only condition we need:

\begin{theorem}\label{thm:Toffoli}
The Toffoli gate and {\em any} basis-changing single-qubit
real gate are universal for quantum computing.
\end{theorem}

The beautiful Gottesman-Knill Theorem~\cite{Gottesman98}
implies that any circuit involving \txtCNOT\ 
and \txtHadamard\ only can be simulated efficiently by a classical circuit. 
It is natural to ask what if \txtHadamard\ is replaced by some other
gate. We know that if this replacement $R$ is a rotation by an irrational
(in degrees) angle, then $R$ itself generates a dense subset of
all rotations, and thus is universal together with \txtCNOT,
by Barenco et al.~\cite{Barenco+95}. What if the replacement
is a rotation of rational angles?
We show that
\txtHadamard\ and its alike are the only exceptions
for a basis-changing single-qubit real gate,
in conjunction with \txtCNOT, to be universal.

\begin{theorem}\label{thm:CNOT}
Let $T$ be a single-qubit real gate and  $T^2$ does
not preserve the computational basis. Then $\set{\txtCNOT,
T}$ is universal for quantum computing.
\end{theorem} 

A basis is said to be complete if it generates a dense
subgroup of $U(k)$ modular a phase, or $O(k)$
for some $k\ge 2$. We actually prove that each of the
two bases in the above theorems gives rise to
a complete basis. By the fundamental theorem of 
Kitaev~\cite{Kitaev97} and Solovay~\cite{Solovay95},
any complete basis can \emph{efficiently} approximate 
any gate (modular a phase),  
or real gate if the basis is real. Therefore,
any real gate can be approximated
with precision $\epsilon$ using $polylog(\inverse{\epsilon})$
gates from either basis, and any circuit
over any basis can be simulated with little blow-up
in the size.

We also provide an alternative prove for
theorem~\ref{thm:Toffoli} by directly constructing
the approximation circuit for an arbitrary real single-qubit
gate, instead of using Kitaev-Solovay theorem.
The drawback of this construction is that the approximation
is polynomial in $\inverse{\epsilon}$; however, it is
conceptually simpler, and uses some new idea that does
not seem to have appeared before (for example, in
the approximation for Control-sign-flip).

There is a broader concept of universality based on 
computations on encoded qubits, that is, fault-tolerant 
quantum computing. We do not discuss this
type of computation, an interested reader is
referred to the survey of Preskill~\cite{Preskill97}.
For a more detailed reference to related works,
refer to the book of Nielsen and Chuang~\cite{NielsenC00}.

\section{Preliminary}
Denote the set $\set{1, 2, \cdots, n}$ by $[n]$.
We will mostly follow the notations and definitions from
the book by Kitaev, Shen, and Vyalyi~\cite{KitaevSV02}.

The (pure) state of a quantum system is a unit vector
in its state space.
The state space of one quantum bit, or qubit, is the
two dimensional complex Hilbert space, denoted by
$\Hspace$. A prechosen orthonormal basis
of $\Hspace$ is called the computational basis
and is denoted by $\set{\ket{0}, \ket{1}}$. 
The state space of a set of $n$ qubits
is the tensor product of the state space of each
qubit, and the computational basis
is denoted
by \[\set{\ket{b} = \ket{b_1}\otimes\ket{b_2}\otimes\cdots\otimes
\ket{b_n} : b=b_1b_2\cdots{}b_n\in\bstr{n}}.\]
A gate is a unitary operator $\op{U}\in \opSet{U}(\tensor{\Hspace}{r})$,
for some integer $r>0$. For an ordered subset $A$
of a set of $n$ qubits,
we write $\op{U}[A]$ to denote applying $\op{U}$ to
the state space of those qubits.
A set of gates is also called a {\em basis}.
A \emph{quantum circuit} over a basis $G$, or
a $G$-circuit, on $n$ qubits and of size $m$ 
is a sequence $\op{U}_1[A_1],\;\op{U}_2[A_2],
\;\cdots,\ \op{U}_m[A_m]$, where each 
$\op{U}_i\in G$ and $A_i\subseteq[n]$. Sometimes we use the same 
notation for a circuit and for
the unitary operator that it defines.
The following definition 
generalizes the definition in \cite{KitaevSV02}.

\begin{definition}
The operator $\op{U} : \SHilbert{\otimes r}\rightarrow\SHilbert{\otimes r}$
is approximated by the operator 
$\tilde{\op{U}} : \SHilbert{\otimes N}\rightarrow\SHilbert{\otimes N}$
using the ancilla state $\ket{\Psi}\in\SHilbert{\otimes N-r}$ if,
for arbitrary vector $\ket{\xi}\in \tensor{\Hspace}{r}$,
\[\norm{ \tilde{\op{U}}(\ket{\xi}\otimes\ket{\Psi}) -
\op{U}\ket{\xi}\otimes\ket{\Psi}} \;\le\; \epsilon\norm{\ket{\xi}}.\]
\end{definition}

Let $G$ be a basis. A {\em $G$-ancilla state}, or an ancilla
state when $G$ is understood, of $\ell$ qubits
is a state $\op{A}\ket{b}$, for some
$G$-circuit $\op{A}$ and some $b\in\bstr{\ell}$.
A basis $G$ is said to be {\em universal for quantum
computing} if any gate (modular a phase), or any
real gate when each gate in $G$ is real,
can be approximated with arbitrary precisions
by $G$-circuits using $G$-ancillae.
By a phase, we mean the set $\set{\exp(i\alpha) : \alpha\in\mathbb{R}}$.
The basis is said to be complete if
it generates a dense subgroup of $U(k)$ modular a phase,
or $O(k)$ when its real for some
$k\ge 2$. A complete basis is clearly universal.

Now we introduce the standard notations for
some gates we shall use later.
Denote the identity operator on $\Hspace$ by $\opId$. 
We often identify a unitary operator by its action on the 
computational basis. The Pauli operators
$\sigma^x$ and $\sigma^z$, and the {\em Hadamard
gate} $\opH$ are
\begin{displaymath}
\sigma^x \define \left(
	\begin{array}{cc}
	0&1\\1&0
	\end{array} \right),
\quad
\sigma^z \define\left(
	\begin{array}{cc}
	1&0\\0&-1
	\end{array} \right),
\quad
\opH\define\frac{1}{\sqrt{2}}\left(
	\begin{array}{cc}
	1&1\\1&-1\end{array}\right).
\end{displaymath}

If $\op{U}$ is a gate on $r$ qubits,
for some $r\ge0$ (when $r=0$, $\op{U}$ is a phase factor),
$\Ctrl[k]{\op{U}}$ is the gate on $k+r$ qubits
that applies $\op{U}$ to the last $r$ qubits
if and only if the first $k$ qubits are in 
$\tensor{\ket{1}}{k}$.
The superscript $k$ is omitted if $k=1$.
Changing the control condition to be
$\tensor{\ket{0}}{k}$, we obtain $\CtrlN[k]{\op{U}}$.
The {\txtToffoli\ gate} is $\Lambda^2{(\sigma^x)}$,
and {\txtCNOT} is $\Lambda{(\sigma^x)}$. Evidently
the latter can be realized  by the former.
From now on we only consider real gates. As in the
previous section, a gate $g$ is said to be {\emph
basis-changing} if it does not preserve
the computational basis. 
\section{Completeness proofs}
In this section we prove the following theorems,
from which Theorem~\ref{thm:CNOT} and 
Theorem~\ref{thm:Toffoli} follow immediately.
\begin{theorem}\label{thm:cnot}
Let $S$ be any single-qubit real gate that is
basis-changing after squaring. Then 
$\set{\txtCNOT, S}$ is complete.
\end{theorem}
\begin{theorem}\label{thm:toff}
The set $\set{\opToffoli, \opHadamard}$ is
complete.
\end{theorem}

We need the following two lemmas, which fortunately
have been proved.

\begin{lemma}[W{\l}odarski~\cite{Wlodarski69}]
\label{lm:angle} If $\alpha$ is not an integer multiple
of $\pi/4$,
and $\cos\beta = \cos^2 \alpha$, then either
$\alpha$ or $\beta$ is an irrational multiple of
$\pi$.
\end{lemma}

\begin{lemma}[Kitaev~\cite{Kitaev97}]\label{lm:group}
Let $\mathcal{M}$ be a Hilbert space of dimension $\ge3$,
$\ket{\xi}\in\mathcal{M}$ a unit vector, and $H\subset SO(\mathcal{M})$
be the stabilizer of the subspace $\mathbb{R}(\ket{\xi})$.
If $\op{V}\in O(\mathcal{M})$ does not preserve $\mathbb{R}(\ket{\xi})$,
$H\bigcup \op{V}^{-1} H \op{V}$ generates a dense
subgroup of $SO(\mathcal{M})$.
\end{lemma}

\begin{proofof}{Theorem~\ref{thm:cnot}}
Define
\[ \op{U}\define \left(\op{S}\otimes\op{S} \cdot \opCNOT[1, 2]\right)^2.\] 
It suffices to prove that $\op{U}$ and
$\opCNOT$ generate a dense subgroup of $SO(4)$.
Without loss of generality, we assume that
$\op{U}$ is a rotation by an angle $\theta$,
the other case can be proved similarly. Then by
the assumption, $\theta$ is not an integer
multiple of $\pi/4$.

Direct calculation shows that $\op{U}$
has eigenvalues $\set{1, 1, \exp(\pm i\alpha)}$,
where \[\alpha=2\arccos\cos^2\theta.\]
The two eigenvectors with eigenvalue $1$ are
\[\ket{\xi_1}\define\frac{1}{2}(\ket{00}-\ket{01}+\ket{10}+\ket{11}),\]
and
\[\ket{\xi_2}\define\frac{\sin\theta}{\sqrt{2}}
(-\ket{0}+\ket{1}) + \frac{\cos\theta}{\sqrt{2}}
(\ket{0}-\ket{1}).\]
Let $\set{\ket{\xi_i}: i\in[4]}$ be a set of
orthonormal vectors.

By Lemma~\ref{lm:angle},
$\alpha$ is incommensurate with $\pi$, therefore,
$\op{U}$ generates
a dense subgroup of $H_1\define SO(\Span\{\ket{\xi_3}, \ket{\xi_4}\})$.
Note that $\opCNOT[1, 2]$ preserve $\ket{\xi_1}$,
but not $\Span\{\ket{\xi_2}\}$. Therefore, by
Lemma~\ref{lm:group}, the set
\[H_1\bigcup \opCNOT[1, 2]\ H_1\ \opCNOT[1, 2]\]
generates a dense subgroup of \[SO(\Span\{\ket{\xi_i}: i=2, 3, 4\})=:H_2,\]
thus so does $\set{\op{U}, \opCNOT[1, 2]}$.
Finally, observe that $\opCNOT[2, 1]$
does not preserve $\Span\{\ket{\xi_1}\}$, therefore,
apply Lemma~\ref{lm:group} again we conclude
that $\set{\op{U}, \opCNOT[1, 2], \opCNOT[2, 1]}$ 
generates a dense subgroup of $SO(4)$.
\end{proofof}

\begin{proofof}{Theorem~\ref{thm:toff}}
Define
\[\op{U}\define \left( \opHadamard\otimes\opHadamard\otimes\opHadamard
\cdot \opToffoli[1,2,3] \right)^2.\]
Direct calculation shows that $\op{U}$
has eigenvalue $1$ with multiplicity $6$,
and the other two eigenvalues 
$\lambda_{\pm}\define\exp(\pm i\alpha)$,
where $\alpha=\pi - \arccos\frac{3}{4}$.
Since $\lambda_{\pm}$ are roots of
the irreducible polynomial
\[ \lambda^2 -\frac{3}{2}\lambda + 1,\]
which is not integral, therefore
$\lambda_{\pm}$ are not algebraic integers. Thus
$\alpha$ is incommensurate with $\pi$,
which implies that $\op{U}$ generates
a dense subgroup of the rotations
over the corresponding eigenspace (denote the
eigenvectors by $\ket{\xi_7}$ and $\ket{\xi_8}$).

By direct calculation, the eigenvectors
correspond to eigenvalue $1$ are:
\[\set{\ket{000},
\ket{010}, \ket{100}, \ket{001}+\ket{011},
\ket{101}+\ket{110}+\ket{111},
\ket{011}-\ket{101}}.\]
Label the above eigenvectors by
$\ket{\xi_i}$, $i\in[6]$.
It is easy to verify that
each $\op{U}_i$, $i\in[6]$, constructed below
preserves
$\set{\ket{\xi_j} : 1\le j<i}$, but
not $\Span\{\ket{\xi_i}\}$. 
\begin{eqnarray*}
\op{U}_1\define\opId\otimes\opId\otimes\opHadamard,\quad&&\quad
\op{U}_2\define \op{U}_1 \cdot \opToffoli[2, 3, 1] \cdot \op{U}_1,\\
\op{U}_3\define\op{U}_1\cdot\opToffoli[1,3,2]\cdot\op{U}_1,\quad&&\quad
\op{U}_4\define\opToffoli[2,3,1],\\
\op{U}_5\define\op{U}_1\cdot \opToffoli[2,3,1]\cdot\op{U}_1,\quad&&\quad
\op{U}_6\define\opToffoli[1,3,2].
\end{eqnarray*}
Applying
Lemma~\ref{lm:group} several times, we see that
$\set{\op{U}, \op{U}_i, \op{U}_{i+1}, \cdots,
\op{U}_6}$ generates a dense subgroup
of $\Span\{\ket{\xi_j}: i\le j\le 8\}$.
Thus $\set{\opToffoli, \opHadamard}$ generates a dense subgroup
of $SO(8)$.
We leave the details for the interested readers.
\end{proofof}

\section{Alternative proof for Theorem~\ref{thm:Toffoli}}
Fix an arbitrary basis-changing real single-qubit
gate $\op{S}$, and the basis 
\[\RealB\define\set{\op{S}, \opToffoli}.\]
In this section we give an explicit construction
to approximate an arbitrary real gate using the
basis $\RealB$. 
Due to the following result
by Barenco et al.~\cite{Barenco+95},
we need only consider approximating single-qubit
real gates:

\begin{proposition}[Barenco et al.~\cite{Barenco+95}]\label{prop:divi}
Any gate on $r$ qubits can be 
realized by $O(r^24^r)$ \txtCNOT{}  and single-qubit
gates.
\end{proposition}

Fix an arbitrary
single-qubit gate $\op{W}$ that we would like to approximate.
Without loss of generality, we can assume
that $\op{S}$ and $\op{W}$ are rotations, 
for otherwise $\sigma^x\op{S}$ and $\sigma^x\op{W}$ are. 
For any $\beta\in[0, 2\pi)$,
define 
\begin{displaymath}
\ket{\phi_\beta} \define
         \cos\beta\ket{0} + \sin\beta\ket{1},
\qquad\textrm{and,}\qquad
 \op{U}_\beta \define 
	\left( \begin{array}{ccc}
		\cos\beta & -\sin\beta\\
		\sin\beta & \cos\beta
		\end{array}
	\right).
\end{displaymath}
Let $\theta, \alpha\in[0, 2\pi)$,
and $\theta$ not an integral multiple of $\pi/2$, be such that
$\op{S}\equiv\OpU$ and $\op{W}\equiv\OpUA$.
The following proposition can be easily checked. 

\begin{proposition}\label{prop:OpUA}
Let $\OpWHA$ be a gate 
on $k+1$ qubits such that
$\OpWHA\tensor{\ket{0}}{k+1}= \SHA\otimes\tensor{\ket{0}}{k}$.
With
\begin{equation}
\label{eqn:OpUA}
 \OpWA\define \OpWHA\;(-\CtrlN[k+1]{-1})\;
 \OpWHA^{\dagger}\;\sigma^z[1],
\end{equation}
for any vector $\ket{\xi}\in{\Hspace}$,
\begin{equation}\label{eqn:OpUA2}
\OpUA\ket{\xi}\otimes\tensor{\ket{0}}{k} = \OpWA
(\ket{\xi}\otimes\tensor{\ket{0}}{k}).
\end{equation}
\end{proposition}

Clearly $\CtrlN[k+1]{-1}$ can be realized
by $\opToffoli$ and $\sigma^z$. Therefore, to approximate
$\OpUA$, it suffices to approximate $\sigma^z$ and
$\OpWHA$, which we will show
in the following subsections.
Define the constants
\[ \ConstSin\define 1/\log\frac{1}{\cos^4\theta + \sin^4\theta},
\quad\textrm{and,}\qquad
\ConstCos\define 1/\log\frac{1}{\cos^2\theta}.\]

\subsection{Approximating $\sigma^{z}$}\label{subsec:AppSigmaZ}
If $\theta$ is a multiple of $\pi/4$, say $\theta=\pi/4$,
then we can easily do a sign-flip by
applying a bit-flip on $\OpU\ket{1} = 
\frac{1}{-\sqrt{2}}\ket{0} + \frac{1}{\sqrt{2}}\ket{1}$.
But for a general $\theta$,
$\op{U}_\theta\ket{1} = -\sin\theta\ket{0}+\cos\theta\ket{1}$
is ``biased''. Immediately comes into mind
is the well-known idea of von Neumann on how to
approximate a fair coin by tossing a sequence of
coins of identical bias\footnote{That is,
toss two coins, declare ``$0$'' if the outcomes
are ``$01$'', declare  ``$1$'' if the outcomes
are ``$10$'', and continue the process otherwise.}.
To illustrate the idea, consider
\[ \OpU\ket{0}\otimes\OpU\ket{1} =
\sin\theta\cos\theta(\ket{11} - \ket{00}) \;+\;
\cos^2\theta\ket{01} - \sin^2\theta\ket{10}.\]
If we switch $\ket{00}$ and $\ket{11}$ and 
leave the other two base vectors unchanged, the first term
on the right-hand side changes the sign, while the
remaining two terms are unchanged. While we
continue tossing pairs of ``quantum coins''
and do the $\ket{00}$-and-$\ket{11}$
switch, we approximate the sign-flip very quickly.
The state defined below will serve the role of 
$\frac{1}{\sqrt{2}}\ket{1} - \frac{1}{\sqrt{2}}\ket{0}$.

\begin{definition} For any integer $k\ge0$,
the \emph{phase ancilla} of size $k$ is the state
\[\PhaseAncila\define (\OpU\ket{0}\otimes\OpU\ket{1})^{\otimes{k}}.\]
\end{definition}
Clearly $\PhaseAncila$ can be prepared
from $\ket{0}^{\otimes 2k}$ by a $\RealB$-circuit
of size $O(k)$.

\begin{lemma}\label{lm:sigmaz}
The operator $\sigma^z$ can be approximated
with precision $\epsilon$, for any $\epsilon>0$, 
by a $\RealB$-circuit of size $O(k)$, using
the phase ancilla $\PhaseAncila$, for some integer
$k=\ub{\ConstSin\ConstLIErr}$.
\end{lemma}

\begin{proof}
Let $k$ be an integer to be determined later. 
The following algorithm is a description of
a circuit approximating $\sigma^z$ using $\PhaseAncila$.

\parbox[c]{5.7in}{
\begin{algorithm}\label{alg:AppSigmaZ}
A $\RealB$-circuit $\tilde{\sigma}^z$ approximating $\sigma^z$
using the phase ancilla $\PhaseAncila$.\\
Let $\ket{b_0}\otimes\ket{b}$ be a computational base vector,
where $b_0\in\bstr{}$ is the qubit to which $\sigma^z$ is to applied,
and $b=b_1 b'_1 b_2 b'_2 \cdots b_k b'_k \in\bstr{2k}$
are the ancilla qubits.
Condition on $b_0$ (that is, if $b_0=0$, do nothing, otherwise
do the following),
\begin{enumerate}
\addtolength{\itemsep}{-11pt}
\addtolength{\topsep}{-11pt}
\item[]{Case 1:} There is no $i$ such that $b_i\oplus b'_i=0$, do nothing.
\item[]{Case 2:} Let $i$ be the smallest index such that $b_i\oplus b'_i=0$,
flip $b_i$ and $b'_i$.
\end{enumerate}
\end{algorithm}
}

Clearly the above algorithm can be carried out
by $O(k)$ applications of Toffoli.
Fix an arbitrary unit vector $\ket{\xi}\in\Hspace$.
Since neither $\sigma^z$ nor $\tilde{\sigma}^z$ changes  
$\ket{0}\bra{0} (\ket{\xi}\otimes\PhaseAncila)$, 
\begin{equation}\label{eqn:AppSigmaZ}
 \norm{ \sigma^z \ket{\xi}\otimes \ket{\Phi_k} - 
\OpAppSigmaZ (\ket{\xi}\otimes\ket{\Phi_k})}
\;\le\;  \norm{-\ket{1}\otimes\PhaseAncila -
		\OpAppSigmaZ (\ket{1}\otimes\PhaseAncila)}.
\end{equation}
Let $\ket{\Phi^+_k}$ ($\ket{\Phi^-_k}$)
be the projection of $\PhaseAncila$ to the subspace
spanned by the base vectors satisfying Case (1) (Case (2)), 
it is easy to prove by induction that
\[ \tilde{\sigma}^z (\ket{1}\otimes\ket{\Phi^+_k}) = 
\ket{1}\otimes\ket{\Phi^+_k},\quad \textrm{and},\quad
\tilde{\sigma}^z (\ket{1}\otimes\ket{\Phi^-_k}) = - 
\ket{1}\otimes\ket{\Phi^-_k}.\]
Furthermore,
\[ \norm{\ket{\Phi^+_k}} = \left(\cos^4\theta + \sin^4\theta\right)^{k/2}.\]
Therefore, the left-hand side of Equation~\ref{eqn:AppSigmaZ}
is upper bounded by
\[2\norm{\ket{\Phi^+_k}} = 2\left(\cos^4\theta +\sin^4\theta\right)^{k/2}.\]
Since $\theta$ is not a multiple of $\pi/2$, the right-hand
side is $<1$.
Thus choosing $k = O(\ConstSin\ConstLIErr)$, the right-hand
side in the above can be made $\le\epsilon$.
\end{proof}

\subsection{Creating $\HalfAlpha$}\label{subsec:halfalpha}
We would like to construct a circuit
that maps $\ket{0}\otimes\sCopies{\ket{0}}{k}$ to 
a state close to $\SHA\otimes\sCopies{\ket{0}}{k}$. 
The main idea is to 
create a ``logical'' $\SHA$:
\begin{equation}\label{eqn:logHA}
\LogHA \define 
\cos\frac{\alpha}{2}\ket{\hat{0}} + \sin\frac{\alpha}{2}\ket{\hat{1}},
\end{equation}
where $\LogX$ and $\LogY$ are two orthonormal vectors 
in a larger space spanned by ancillae, and then
undo the encoding to come  back
to the computational basis.
To create $\LogHA$, we first create a state almost
orthogonal to $\LogX$, and then apply Grover's algorithm~\cite{Grover96}
to rotate this state toward $\LogHA$.
Define the operator $\OpT$ on $2$ qubits as
\begin{equation}
\label{eqn:OpT}
\OpT\define \OpUI[1]\;\opCNOT[1, 2]\; \OpU[1].
\end{equation}
Since for any $\beta$, 
$\op{U}_{-\beta} = \sigma^x \op{U}_\beta \sigma^x$,
$\OpT$ and $\Ctrl{\OpT}$ can be realized by 
the basis $\RealB$. 
Let \[\RealB_1 \define \set{\opToffoli, \sigma^z,
\OpU, \OpUI, \OpT, \Ctrl{\opT}}.\]

\begin{lemma}\label{lm:AppOpWHA}
For any $\epsilon>0$ there exists a $\RealB_1$-circuit 
$\AppOpWHA$ of size 
$O(\ConstCos\ConstIErr\ConstLIErr)$
that uses $O(\ConstCos\ConstLIErr)$
ancillae and satisfies
\[  \norm{ \AppOpWHA \ket{0}^{\otimes k+1} - 
  \SHA\otimes\ket{0}^{\otimes k}} \le \epsilon.\]
\end{lemma}

\begin{proof}
Figure~\ref{fig:OpUA} illustrates our proof.
Let $k>0$ be an integer to be specified later. Define
\[\LogX \define\ket{0}^{\otimes 2k},\qquad
\AppY\define \OpT^{\otimes k} \ket{\hat{0}},\qquad
\textrm{and,}\qquad
\gamma\define\arcsin\left(\cos^{2k}\theta\right).\]
Notice that $\pi/2-\gamma$ is the angle between
$\LogX$ and $\AppY$, and $0<\gamma<\pi/2$,
since $\sin\gamma = \inner{\hat{0}}{\tilde{1}}$.
Let $S$ be the plane spanned by $\LogX$ and $\AppY$
Let $\LogY$ be the unit vector perpendicular to 
$\LogX$ in $S$ and the angle between $\LogY$ and $\AppY$ is 
$\gamma$.

Observe that on $S$ we can do the reflection along
$\LogY$ and the reflection along $\AppY$. The former
is simply $\CtrlN[2k]{\sigma^z}$,
which can be implemented using $\opToffoli$ and $\sigma^z$.
Since $\OpT^{-1} = \OpT$, the reflection along
$\AppY$ is
\[\op{R}\define\tensor{\OpT}{k}\;(-\ReflectY)\;\tensor{\OpT}{k}.\]
Without loss of generality we can assume
$\alpha/2<\pi/2$; otherwise we will rotate $\AppY$
close to $\ReflectY\LogHA$ and then apply
$\ReflectY$.
Choose $k$ sufficiently large so that
$\gamma<\pi/2-\alpha/2$. Now we can
apply Grover's algorithm to 
rotate $\AppY$ to a state very close to $\LogHA$.
After that we do a ``controlled-roll-back'' to
map $\LogY$ (approximately) to $\ket{1}^k$ and does not
change $\LogX$. This will give us an approximation of
$\SHA$ in the state space of the controlling qubit. 
The algorithm is as follows.
Let $T$ be the integer such that $|\pi/2- (2T+1)\gamma - \alpha/2| < \gamma$.
Then $T = O(1/\gamma)$.

\begin{figure}
\begin{center}
\setlength{\unitlength}{0.00083333in}
\begingroup\makeatletter\ifx\SetFigFont\undefined%
\gdef\SetFigFont#1#2#3#4#5{%
  \reset@font\fontsize{#1}{#2pt}%
  \fontfamily{#3}\fontseries{#4}\fontshape{#5}%
  \selectfont}%
\fi\endgroup%
{\renewcommand{\dashlinestretch}{30}
\begin{picture}(4176,3172)(0,-10)
\put(1800.000,1859.500){\arc{604.669}{4.1932}{6.4075}}
\blacken\path(2060.267,1939.138)(2100.000,1822.000)(2120.066,1944.055)(2060.267,1939.138)
\put(900.000,2497.000){\arc{335.410}{2.6779}{5.8195}}
\blacken\path(940.701,2629.911)(1050.000,2572.000)(980.812,2674.533)(940.701,2629.911)
\put(1875.000,847.000){\arc{474.342}{4.3906}{7.5322}}
\blacken\path(1917.929,1109.320)(1800.000,1072.000)(1921.617,1049.433)(1917.929,1109.320)
\blacken\path(2033.796,712.985)(1950.000,622.000)(2066.754,662.847)(2033.796,712.985)
\thicklines
\path(150,622)(3600,622)
\blacken\path(3360.000,562.000)(3600.000,622.000)(3360.000,682.000)(3360.000,562.000)
\path(900,22)(900,3022)
\blacken\path(960.000,2782.000)(900.000,3022.000)(840.000,2782.000)(960.000,2782.000)
\path(900,622)(3150,1672)
\blacken\path(2957.889,1516.136)(3150.000,1672.000)(2907.143,1624.878)(2957.889,1516.136)
\path(900,622)(1200,3022)
\blacken\path(1229.768,2776.411)(1200.000,3022.000)(1110.695,2791.295)(1229.768,2776.411)
\thinlines
\dashline{60.000}(900,622)(1800,2872)
\blacken\path(1783.287,2749.441)(1800.000,2872.000)(1727.579,2771.725)(1783.287,2749.441)
\dashline{60.000}(900,622)(150,2872)
\blacken\path(216.408,2767.645)(150.000,2872.000)(159.487,2748.671)(216.408,2767.645)
\dashline{60.000}(900,622)(2250,2572)
\blacken\path(2206.361,2456.261)(2250.000,2572.000)(2157.029,2490.413)(2206.361,2456.261)
\thicklines
\dashline{90.000}(900,622)(3000,1822)
\blacken\path(2821.390,1650.832)(3000.000,1822.000)(2761.853,1755.021)(2821.390,1650.832)
\put(3900,472){\makebox(0,0)[lb]{\smash{{{\SetFigFont{12}{14.4}{\rmdefault}{\mddefault}{\updefault}$\LogX$}}}}}
\put(3300,1522){\makebox(0,0)[lb]{\smash{{{\SetFigFont{12}{14.4}{\rmdefault}{\mddefault}{\updefault}$\LogHA$}}}}}
\put(2400,2572){\makebox(0,0)[lb]{\smash{{{\SetFigFont{12}{14.4}{\rmdefault}{\mddefault}{\updefault}$\ket{\phi_3}$}}}}}
\put(1950,2872){\makebox(0,0)[lb]{\smash{{{\SetFigFont{12}{14.4}{\rmdefault}{\mddefault}{\updefault}$\ket{\phi_1}$}}}}}
\put(1200,3022){\makebox(0,0)[lb]{\smash{{{\SetFigFont{12}{14.4}{\rmdefault}{\mddefault}{\updefault}$\AppY$}}}}}
\put(750,3022){\makebox(0,0)[lb]{\smash{{{\SetFigFont{12}{14.4}{\rmdefault}{\mddefault}{\updefault}$\LogY$}}}}}
\put(0,2872){\makebox(0,0)[lb]{\smash{{{\SetFigFont{12}{14.4}{\rmdefault}{\mddefault}{\updefault}$\ket{\phi_2}$}}}}}
\put(450,2272){\makebox(0,0)[lb]{\smash{{{\SetFigFont{12}{14.4}{\rmdefault}{\mddefault}{\updefault}$\gamma$}}}}}
\put(2250,922){\makebox(0,0)[lb]{\smash{{{\SetFigFont{12}{14.4}{\rmdefault}{\mddefault}{\updefault}$\alpha/2$}}}}}
\put(3150,1822){\makebox(0,0)[lb]{\smash{{{\SetFigFont{12}{14.4}{\rmdefault}{\mddefault}{\updefault}$\Mixed$}}}}}
\end{picture}
}
\caption{Creating an approximate $\LogHA$. In one 
iteration in Grover's algorithm, $\ket{\phi_1}\rightarrow\ket{\phi_2}\rightarrow
\ket{\phi_3}$.}
\label{fig:OpUA}
\end{center}
\end{figure}

\parbox[c]{5.7in}{
\begin{algorithm}\label{alg:AppWHA}
A $\RealB_1$-circuit $\AppOpWHA$ that maps $\ket{0}\otimes\tensor{\ket{0}}{2k}$
to a state close to $\SHA\otimes\tensor{\ket{0}}{2k}$.
\begin{enumerate}
\addtolength{\itemsep}{-11pt}
\addtolength{\topsep}{-11pt}
\item Apply $\op{I}\otimes\OpTK$.
\item (Grover's algorithm) 
Apply $\left(\op{R}\;\ReflectY\right)^{T}$.
\item (Sub-circuit $\op{A}_3$) For a computational base vector $\ket{b}$ of
the ancillae, if $\ket{b}\ne\LogX$, flip the first bit.
\item (Sub-circuit $\op{A}_4$) Use the first bit as the condition bit, apply
$\Ctrl{\OpTK}$.
\end{enumerate}
\end{algorithm}}

It can be easily verified that
\[\norm{\AppOpWHA (\ket{0}\otimes\sCopies{\ket{0}}{2k}) 
- \SHA\otimes\sCopies{\ket{0}}{2k}} \;\le\;2\gamma.\]
Setting $\gamma \approx \epsilon/2$,
by direct computation the number of ancillae is
$O(k) = O(\ConstCos\ConstLIErr)$,
and the size of  $\AppOpWHA$ is
$O(k/\gamma) = O(\ConstCos\ConstIErr\ConstLIErr)$.

\end{proof}

\subsection{Approximating $\OpUA$}\label{subsec:OpUA}
Theorem~\ref{thm:Toffoli} is a straightforward corollary of
the following theorem and Proposition~\ref{prop:divi}.
\begin{theorem}\label{thm:OpUA}
For any $\epsilon>0$, the operator $\OpUA$ 
can be approximated with precision $\epsilon$
by a $\RealB$-circuit of size 
$\ub{\ConstSin\cdot\ConstIErr\cdot\ConstLIErr}$
and using $\ub{\ConstSin\cdot\ConstLIErr}$
ancillae.
\end{theorem}

\begin{proof}
We first compose a $\RealB_1$-circuit
that approximates $\OpUA$, according to
Equation~\ref{eqn:OpUA}, \ref{eqn:OpUA2},
 and Algorithm~\ref{alg:AppWHA},
and use $k_1$ (different) ancillae in each call to the latter,
for an integer $k_1$ to be specified later.
Let $\gamma\define\cos^{2k_1}\theta$. Then the precision
is $O(\gamma)$.
After implementing $\OpT$ and $\Ctrl{\OpT}$,
there are in total $O(\ConstIErr)$ uses of
$\sigma^z$. 

Finally we apply Algorithm~\ref{alg:AppSigmaZ}
to approximate each $\sigma^z$ using the {\em same}
phase ancilla $\PhaseAncila[k_2]$ for 
$k_2 = O(1/\gamma^3)$. Let 
$\delta_\theta:=2(\cos^4\theta+\sin^4\theta)^{k_2/2}$
be the error of one call to $\AppSigmaZ$ using exactly
$\PhaseAncila[k_2]$.
Observe that using the same
phase ancilla for
$O(\inverse{\gamma})$ times causes error at most
$1+2+\cdots+O(\inverse{\gamma})-1 = O(\inverse{\gamma^2})$ 
Setting $\delta_\theta = {\gamma^3}$, 
the total error caused by $\AppSigmaZ$ is $O(\gamma)$.
Thus the total error
of the whole circuit is still 
$O(\gamma)$. Setting $\gamma \approx \epsilon$,
$k_1=O(\ConstCos\ConstLIErr) = O(\ConstSin\ConstLIErr)$
and $k_2=O(\ConstSin\ConstLIErr)$. Therefore
the number of ancillae is 
$O(k_1+k_2) = O(\ConstSin\ConstLIErr)$.
The size of the circuit is $O((k_1+k_2)\ConstIErr) = 
O(\ConstSin\ConstIErr\ConstLIErr)$.

\end{proof}
\section{Acknowledgment}
I would like to thank Barbara Terhal, Alexei Kitaev, and Zhenghan Wang
for stimulating and 
helpful discussions, and Ronald de Wolf for valuable comments.


\begin{thebibliography}{10}

\bibitem{Barenco+95}
A.~Barenco, C.~H. Bennett, R.~Cleve, D.~P. DiVincenzo, N.~H. Margolus, P.~W.
  Shor, T.~Sleator, J.~A. Smolin, and H.~Weinfurter.
\newblock Elementary gates for quantum computation.
\newblock {\em Physical Review A}, 52(5):3457--3467, 1995.

\bibitem{BoykinMPRV00}
P.~O. Boykin, T.~Mor, M.~Pulver, V.~Roychowdhury, and F.~Vatan.
\newblock A new universal and fault-tolerant quantum basis.
\newblock {\em Information Processing Letters}, 75(3):101--107, Aug. 2000.

\bibitem{Gottesman98}
D.~Gottesman.
\newblock The Heisenberg representation of quantum computers.
\newblock In S.~P. Corney, R.~Delbourgo, and P.~D. Jarvis, editors, {\em
  Group22: Proceedings of the XXII International Colloquium on Group
  Theoretical Methods in Physics}, pages 32--43, Cambridge, MA, 1999.
  International Press.
\newblock Long version: quant-ph/9807006.

\bibitem{Grover96}
L.~K. Grover.
\newblock A fast quantum mechanical algorithm for database search.
\newblock In {\em Proceedings of the Twenty-Eighth Annual ACM Symposium on the
  Theory of Computing}, pages 212--219, Philadelphia, Pennsylvania, May 1996.

\bibitem{Kitaev97}
A.~Y. Kitaev.
\newblock Quantum computations: Algorithms and error correction.
\newblock {\em RMS: Russian Mathematical Surveys}, 52(6):1191--1249, 1997.

\bibitem{KitaevSV02}
A.~Y. Kitaev, A.~H. Shen, and M.~N. Vyalyi.
\newblock {\em Classical and quantum computation}.
\newblock To appear.

\bibitem{NielsenC00}
M.~A. Nielsen and I.~L. Chuang.
\newblock {\em Quantum Computation and Quantum Information}.
\newblock Cambridge University Press, Cambridge, UK, 2000.

\bibitem{Preskill97}
J.~Preskill.
\newblock Reliable quantum computers.
\newblock 1997.
\newblock quant-ph/9705031.

\bibitem{Solovay95}
R.~Solovay.
\newblock Unpublished manuscript, 1995.

\bibitem{Wlodarski69}
{\L}.~W{\l}odarski.
\newblock On the equation ${\rm cos}\alpha \sb{1}+{\rm cos}\alpha \sb{2} {\rm
  cos}\alpha \sb{3}+{\rm cos}\alpha \sb{4}=0$.
\newblock {\em Ann. Univ. Sci. Budapest. E\"otv\"os Sect. Math.}, 12:147--155,
  1969.

\end{thebibliography}
\end{document}